\documentclass[12pt]{iopart}
\usepackage{verbatim}
\usepackage{graphicx}

\let \IG \includegraphics
\begin{document}

\title[Performance of a 1200\,m long suspended Fabry-Perot cavity]{Performance of a 1200\,m long suspended Fabry-Perot cavity}
\author{A~Freise\footnote[1]{To whom correspondence should be addressed 
(adf@mpq.mpg.de)}\ddag, M~M~Casey\S, S~Gossler\ddag, H~Grote\ddag, G~Heinzel\P, H~L\"uck\ddag\P, D~I~Robertson\S, K~A~Strain\S, H~Ward\S, B~Willke\ddag\P, J~Hough\S\ and K~Danzmann\ddag\P}

\address{\ddag\ Institut f\"ur Atom- und Molek\"ulphysik, 
Universit\"at Hannover, Callinstra\ss e 38, D-30167 Hannover}
\address{\P\ Max-Planck-Institut f\"ur Gravitationsphysik (Albert-Einstein-Institut),
Institut Hannover, Callinstra\ss e 38, D-30167 Hannover}
\address{\S\ Department of Physics and Astronomy, University of Glasgow, Glasgow G12 8QQ, Scotland, United Kingdom}
\begin{abstract}

Using one arm of the Michelson interferometer  
and the power recycling mirror of the interferometric gravitational wave detector
GEO\,600,
we created a Fabry-Perot cavity with a length of 1200\,m. 
The main purpose of this experiment was to gather first experience with the 
main optics, its suspensions and the corresponding control systems. 
The residual displacement of a main mirror is about 150\,nm rms.
By stabilising the length of the 1200\,m long cavity to the pre-stabilised 
laser beam we achieved an error point frequency 
noise of 100\,$\rm \mu Hz /\sqrt{\rm Hz}$ at 100\,Hz Fourier frequency.
In addition we demonstrated the reliable performance of all included
subsystems by several 10-hour-periods of continuous stable operation. 
Thus the full frequency stabilisation scheme for GEO\,600 was 
successfully tested.

\end{abstract}

\submitto{\CQG}

\ead{adf@mpq.mpg.de}

\maketitle

\section{Introduction}

\begin{figure}[htb]
\begin{center}
\IG [scale=.25] {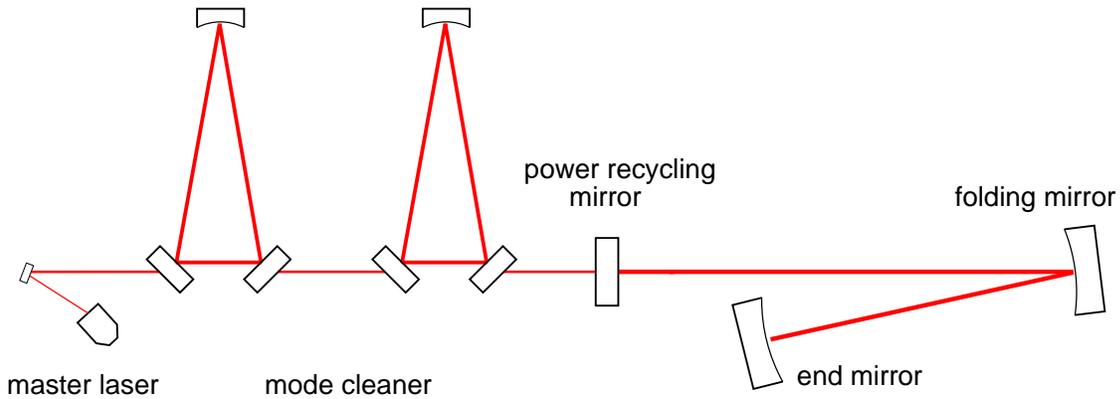} 
\end{center}
\caption{\label{fig:prc_layout}The 1200\,m cavity in January 2001: The light of the
master laser is filtered by two mode cleaners and injected 
into the 1200\,m long cavity formed by the 
power recycling mirror and the two mirrors of the 
folded east arm of GEO\,600.}
\end{figure}

By the beginning of 2001 the installation of the main optics of the 
gravitational wave detector GEO\,600 \cite{GEO600:actuel} has been well under way.
The Nd:YAG laser source and the mode cleaners 
were completely installed. The frequency and length control system 
as well as an automatic alignment system for the mode cleaners were installed.
These systems are operated in a fully automated, remotely controllable 
fashion and provide the 
input beam for an interferometer in the main vacuum system.
In addition, three main mirrors were suspended, the {\it power recycling mirror},
the {\it folding mirror} and the {\it end mirror} of one arm
of the Michelson interferometer (see Figure~\ref{fig:prc_layout}).
These mirrors form a cavity with 2400\,m round trip length, which is  
very similar to the power recycling cavity in a recycled Michelson 
interferometer.
This enabled us to perform a first test of the length and frequency control.
The
cavity is the first large-scale optical system we have
operated in GEO\,600 and the {\it first arm} of the detector.

\section{Optical Layout}
\begin{figure}[htb]
\begin{center}
%\IG [bb= 0 -150 1840 830, scale=.22] {layout1.eps} 
\IG [scale=.22] {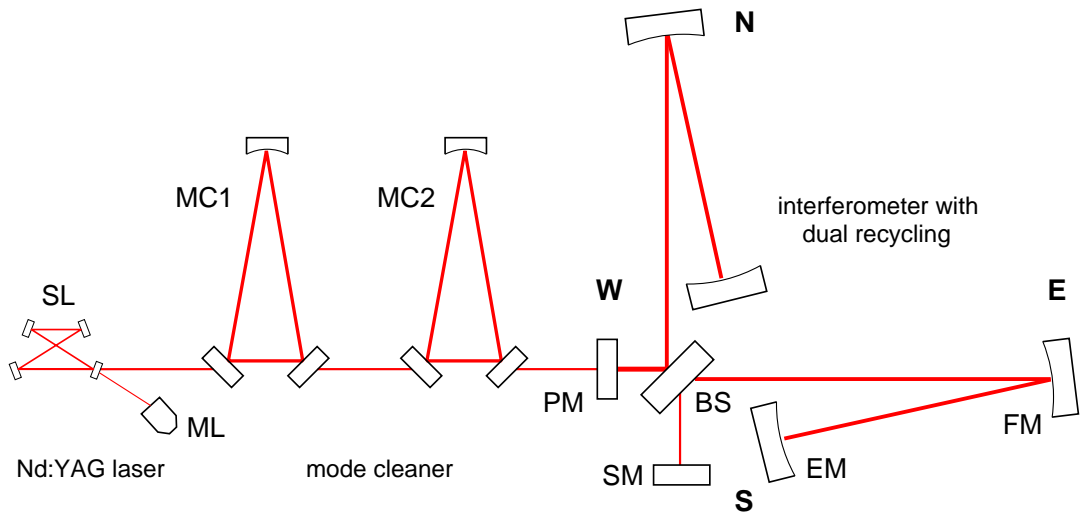} 
\end{center}
\caption{\label{fig:layout1}The optical layout of GEO\,600: The laser consists of a monolithic
master laser plus an injection locked slave laser in a bow tie setup,
the two mode cleaners are suspended 8\,m ring cavities, the main
interferometer is a dual recycled Michelson interferometer with folded arms.}
\end{figure}

The laser system of GEO\,600 is an injection locked master and slave system
with 14\,W output power at 1064\,nm \cite{laser}. After leaving the laser, the light
is passed through two successive mode cleaners \cite{atr:mc}. The mode cleaners 
are
ring cavities with 8\,m round trip length and three mirrors each, which are suspended as double 
pendulums. The measured optical parameters of the two mode cleaners are shown in Table
\ref{tab:mc}.
The main optical instrument in GEO\,600 will be a dual-recycled Michelson 
interferometer 
\cite{ghh:dr,kas:dr}. 
In contrast to other interferometric 
gravitational wave projects \cite{VIRGO:actuel,TAMA:actuel,LIGO:actuel}, GEO\,600 does not 
employ arm cavities 
but folded arms (see Figure~\ref{fig:layout1}). 

The laser light enters the main instrument from the west (W) at the 
{\it power recycling
mirror} (PM) and is split into an east (E) and a north arm (N) 
at the beam splitter (BS). At 
a distance of 600\,m from the beam splitter each arm has a 
{\it folding mirror} (FM)
that directs the beam back (slightly tilted) towards the beam splitter.
The light
hits the {\it end mirror} (EM) located close to the beam splitter
25\,cm above the axis of the injected beam.
In the final optical layout the beams reflected from the two end 
mirrors are superimposed on the beam splitter. The
Michelson interferometer will be held on the so called {\it dark fringe},
where the tiny phase modulation sidebands containing
the gravitational wave signal are directed south (S) towards the {\it
signal recycling mirror} (SM). 
The major part of the light power (the carrier) is reflected
back to the power recycling mirror.  
Thus the power recycling mirror and the
Michelson interferometer form a cavity for the carrier light 
of 2400\,m round trip length, the
{\it power recycling cavity}.

\begin{table}[hbt]
\caption{\label{tab:mc}Measured optical parameters of the GEO\,600 mode cleaners.}

\vspace{.3cm}
\begin{indented}
\item[]\begin{tabular}{@{}clll}
\br
mode cleaner & finesse  & throughput &visibility \\ 
\mr
MC1 &  2700 & 80\% & 94\% \\
MC2 & 1900 & 72\% &92\%\\
\br
\end{tabular}
\end{indented}
\end{table}

All main mirrors are suspended as triple pendulums to give
a very good seismic isolation in the direction of the 
optical axis \cite{mirrors}.
By January 2001 three main mirrors had been installed. The east arm was 
fully equipped with the end mirror and the folding mirror. 
Furthermore, the power recycling mirror was in place. The beam splitter
was left out so that the three mirrors formed a Fabry-Perot cavity
(see Figure~\ref{fig:prc_layout}), which is very similar to the power recycling
cavity of the final detector. For the experiment described here the
master laser (ML), an 1\,W
non-planar ring oscillator (NPRO), was used as light source, with
the 14\,W amplifier (SL) being bypassed.

\section{Laser frequency stabilisation}

The laser frequency stabilisation scheme used in GEO\,600 is unique
because it does not include any rigid reference cavity. Instead cavities
with suspended mirrors are used. 
The suspended cavities provide a very good reference
for frequencies above 50\,Hz because of their excellent
seismic isolation.

\begin{figure}[htb]
\begin{center}
\IG [scale=.27] {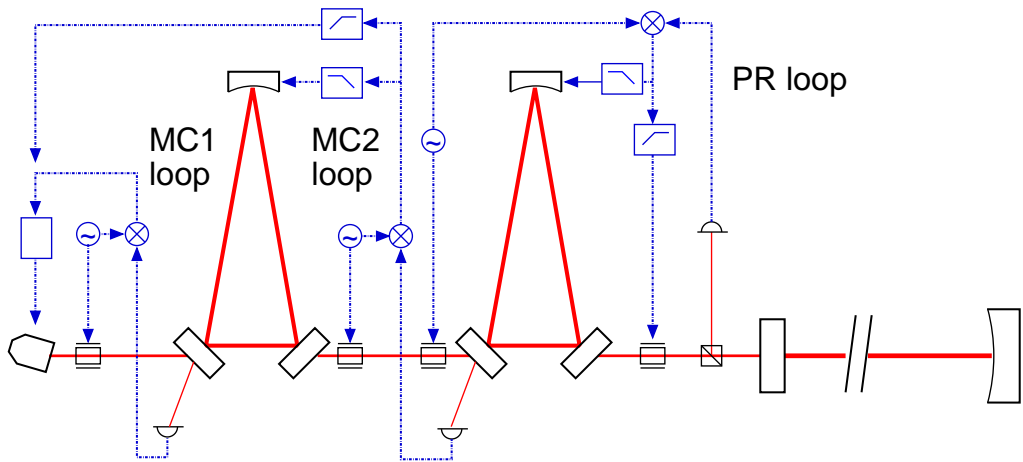} 
\end{center}
\caption{\label{fig:chain}The frequency stabilisation scheme of GEO\,600: The feedback paths 
of the length and frequency control for the
power recycling cavity, the two mode cleaners, and the master laser.}
\end{figure}

The optical systems and the feedback 
systems are shown in Figure~\ref{fig:chain}. 
The frequency of the laser is locked to the first mode cleaner
using a standard Pound-Drever-Hall method (MC1 loop). The feedback is applied
to the frequency modulation input of
the master laser. The bandwidth of this control loop is 100\,kHz.

The next step is to pass the light through the second mode cleaner.
To do so we have to bring the injected light into resonance with the second
mode
cleaner by changing the length of 
the first mode cleaner and thus the laser frequency.
This is done with a feedback control system using  
another Pound-Drever-Hall error signal (MC2 loop). 
The control signal is split and fed back to two actuators: the slow
Fourier components 
($<4$\,kHz) are applied to the length of mode cleaner MC1 via
coil magnet actuators on one of its mirrors, while the fast components  
are injected into the error point of the MC1 loop, i.e. they 
directly act on the master laser frequency (see Figure~\ref{fig:chain}).
The fast feedback path is necessary to achieve a  high servo 
bandwidth (about 25\,kHz) 
and thus a high gain at low Fourier frequencies.  

Finally the pre-stabilised light has to be resonant in the power
recycling cavity. A third control loop (PR loop) 
with yet another Pound-Drever-Hall
scheme is used to control the length of 
mode cleaner MC2 accordingly. 
The feedback of this loop was split into
a `slow' path, which acts on the length of the second 
mode cleaner (by moving
one of its mirrors), and a `fast' path. The fast signal is applied
to an electro-optic modulator in front of the power recycling cavity 
that serves as a fast phase corrector. 
In the
experiment described here the control bandwidth of the PR loop was 40\,kHz.

\begin{figure}[htb]
\begin{center}
\IG [scale=1] {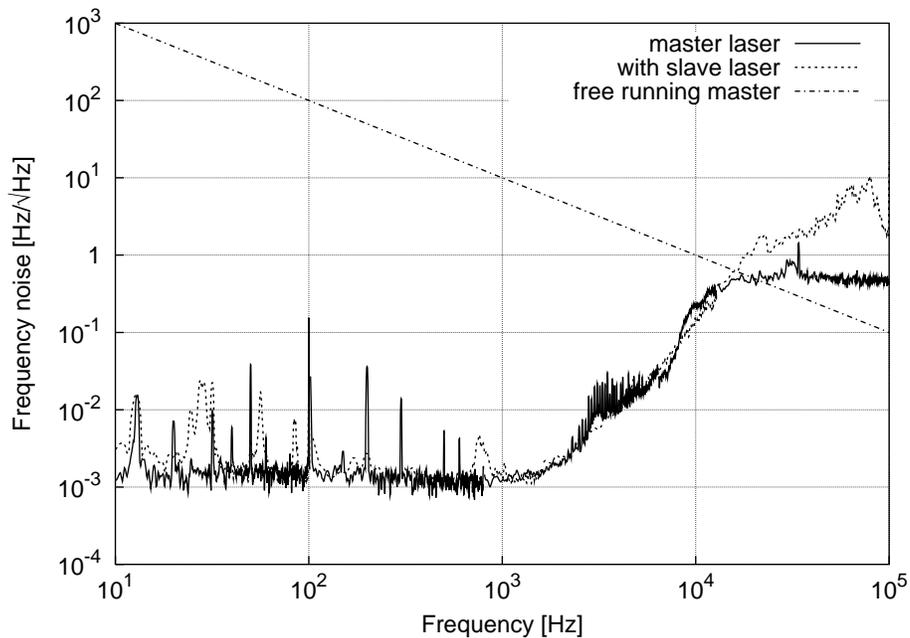} 
\end{center}
\caption{\label{fig:mc_old_new}
The frequency noise at the first mode cleaner (MC1). The in-loop frequency
noise is about 1\,mHz and is not degraded when the high power slave laser
is added.}
\end{figure}

In order to measure the performance of the frequency stabilisation
one can perform two different experiments: a) an in-loop measurement of
the frequency noise (this is done by taking the
error point spectrum of the frequency 
control loop) or b) an out-of-loop measurement of the
residual frequency noise of a stabilised laser with 
respect to an independent
frequency reference.
Figure~\ref{fig:mc_old_new} shows the error point spectrum of the 
MC1 loop that stabilises 
the master laser to the first mode cleaner. 
The frequency noise in the error point is about 
1\,mHz/$\sqrt{\rm Hz}$ below 1\,kHz. In comparison with the typical 
frequency
noise of a free running NPRO one can see the gain of the servo
loop, e.g. 100\,dB at 100\,Hz.

The master and slave laser system can be treated as a black box from the
outside in the sense that the frequency of the slave laser is following
the frequency of the master laser and that thus the slave inherits the frequency
stability of the
master. The injection locking controls the slave laser
frequency with a higher bandwidth 
than the
servo loops that apply feedback to the frequency of the master laser.
Figure~\ref{fig:mc_old_new} also shows a comparison
of the error signal of the MC1 loop 
for a) the master laser only and b) the master plus injection
locked slave. It can be seen that the laser frequency noise 
is very similar in both
cases. The servo electronics of the MC1 loop
were not changed between the two measurements.

\begin{figure}[htb]
\begin{center}
\IG [scale=1] {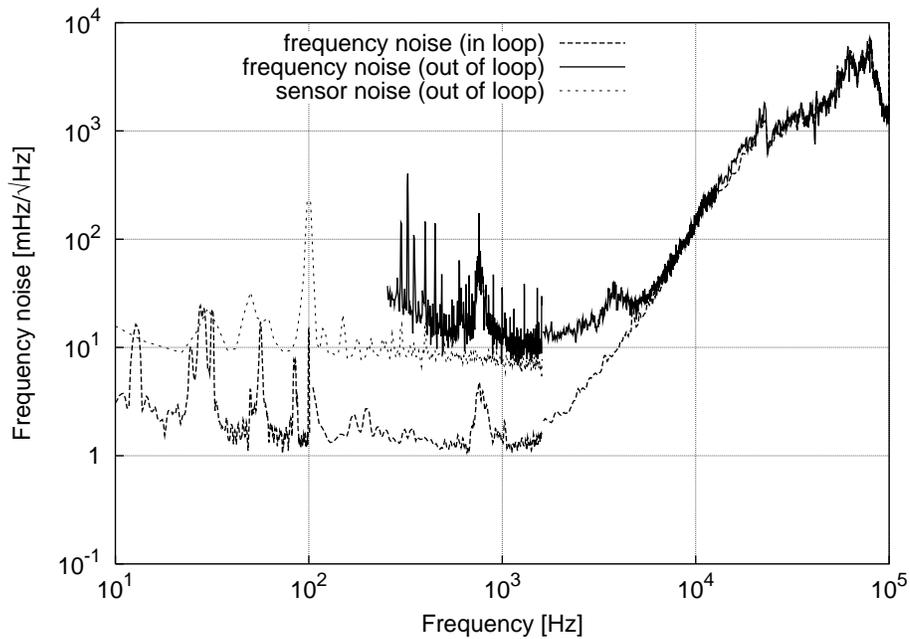} 
\end{center}
\caption{\label{fig:in_out}
Comparison of in-loop and out-of-loop frequency noise 
at mode cleaner MC1. The sensor noise of this 
stabilisation loop is also plotted. Please note that, with the method used
here, the out-of-loop 
frequency noise could
only be measured down to 300\,Hz.}
\end{figure}

With two mode cleaner cavities one can also 
use the second mode cleaner as an independent reference 
(apart from common-mode effects caused by suspending both
mode cleaners from the same mechanical structure in their
vacuum tanks)
to measure the performance of the laser frequency stabilisation
to the first mode cleaner.
This is 
an out-of-loop measurement of the frequency noise and the result is shown 
in Figure~\ref{fig:in_out}.
For this measurement the laser was locked to the first mode cleaner as 
usual.
Then the
first mode cleaner was locked to the
second as described above (MC2 loop) but with a very low servo 
bandwidth (unity gain
frequency $\approx$ 300\,Hz). 
The frequency noise in the error signal
of the MC2 loop then gives the residual frequency noise of the
MC1 loop for Fourier frequencies above the unity gain frequency of the
MC2 loop.

As can be seen in 
Figure~\ref{fig:in_out} the measured out-of-loop
noise floor is about 10\dots30\,mHz/$\sqrt{\rm Hz}$ below 1\,kHz and 
very close to the sensor noise of the MC1 loop,
i.e. the electronic noise of the photo diode. We believe that the shot 
noise limit can be
reached by increasing the light power on the photo diode by a factor of 5.

We also measured the out-of-loop frequency noise of the MC2 loop
with respect to the power recycling cavity. This measurement showed that
the out-of-loop frequency noise of the MC2 loop and that of the MC1 loop
are very similar.
As both mode cleaners are similar cavities with common noise 
sources the frequency noise
can only be improved a little by the second control loop. This is
different for the power recycling cavity: the cavity
has a much smaller bandwidth (400\,Hz in this experiment)
and the length of 1200\,m greatly improves the relative stability. Also
this cavity uses triple pendulum suspensions for its mirrors, so that 
the absolute motion of the mirrors is less than that of the mode cleaner
mirrors with their double pendulum suspension for Fourier frequencies above
10\,Hz.
Thus the power recycling cavity can be used as a stable frequency reference
and by stabilising the mode cleaners and 
the laser light to the power recycling cavity length, the frequency 
noise can be further improved
(see next section).

\begin{figure}[htb]
\begin{center}
\IG [scale=1] {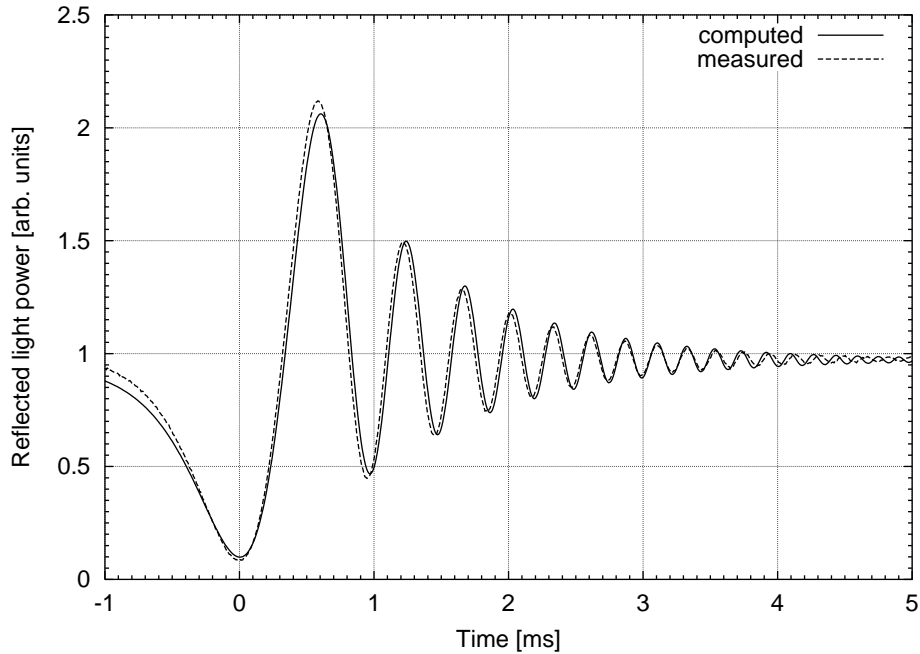} 
\end{center}
\caption{\label{fig:fringe}
Simulated fringe of the power recycling cavity compared to a
measured typical fringe. The simulation gives a finesse of 300 for the power
recycling cavity and a velocity for the relative motion of the mode cleaner 
MC2 mirrors of 20\,nm/s.}
\end{figure}

\section{Performance of the 1200\,m cavity and its length control
system}

The mode cleaner system was completed in December 2000. Both 
cavities were automatically aligned and the lock acquisition of the 
length  
control loops was fully automated. 
The servo systems for the length control use analog
electronic devices. They can be guided by a computer system
via a digital bus \cite{mmc:lv} to
automate lock acquisition of the full optical system. Therefore
in normal operation no human interaction is required. 
We thus had a stable input
beam for the experiment with the 1200\,m cavity.

The three
mirrors of this cavity were without any alignment control and thus had to be 
aligned manually for every experiment. When a reasonably good 
alignment was established one could observe fringes in the light reflected 
from the long cavity. A numerical simulation was used to fit a model function
to a measured time series of the reflected light power of the 1200\,m long cavity (see Figure~\ref{fig:fringe}).
The simulation 
works in the time domain and 
calculates the dynamic changes of the light power inside a  
cavity after an incident laser beam comes into resonance with the cavity. 
In our case the cavity was assumed 
to be rigid
while the frequency of the incoming beam is subject to a sweep (the effect is 
the same as for a cavity with moving mirrors that is illuminated by a light source
with fixed frequency \cite{malik}). This assumption can be made because the relative stability
of the 1200\,m long cavity is much better than that of the mode cleaners, which
function at that time the reference for the laser frequency. The particular fringe
shown in Figure~\ref{fig:fringe} was chosen because it could be modeled with a simple 
linear frequency sweep.

\begin{figure}[htb]
\begin{center}
\IG [bb= 50 50 400 180, clip,scale=1.0] {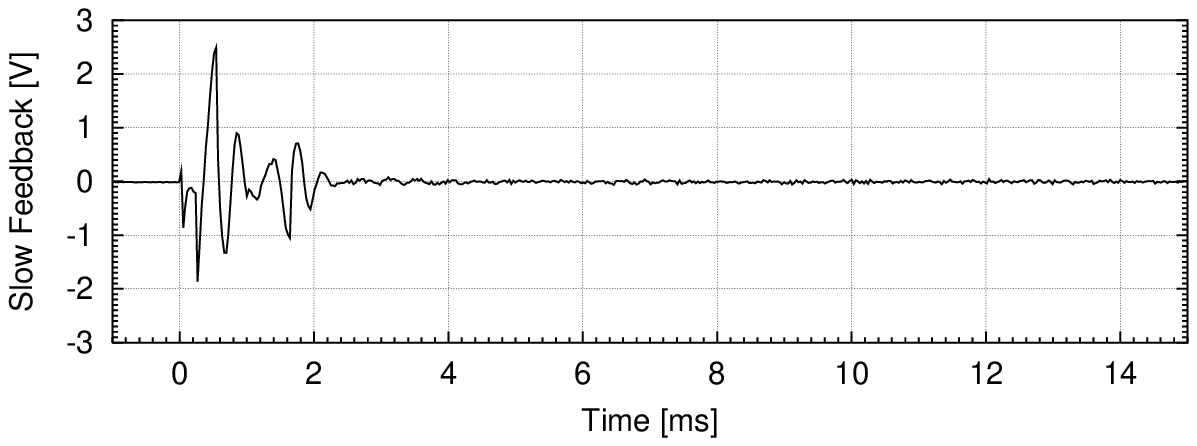} 
\IG [bb= 50 50 400 180, clip,scale=1.0] {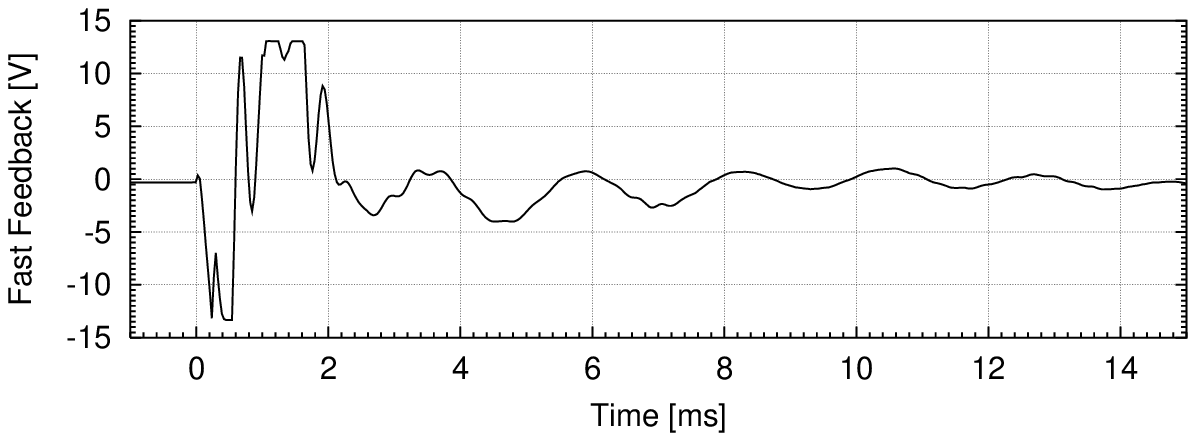} 
\end{center}
\caption{\label{fig:prc_fb}
Time series of feedback signals of the control system for the 1200\,m
long cavity during lock acquisition.}
\end{figure}

The parameters of the fitted model function showed
that the finesse of the cavity (with that particular alignment) was
300 and the speed of the mirrors of the second mode cleaner
was 20\,nm/s. This means that a control
loop locking the incoming light to that cavity had to 
acquire lock in some milliseconds. The low minimum 
of the ringing in Figure~\ref{fig:fringe} shows that the
mode matching is better than 90\%. The finesse of 300 corresponds to
a loss inside the cavity of 0.7\%.

To stabilise the incoming light to the resonance of the 1200\,m long
cavity another feedback control featuring the third Pound-Drever-Hall
scheme was installed (PR loop).
Figure~\ref{fig:prc_fb} shows these feedback signals during lock
acquisition. The servo loop was closed automatically during a fringe
at time zero. It can be seen that the cavity acquired lock in 2\,ms and
stayed in lock while a residual motion of the mirrors damped out quickly.

The automation of the lock acquisition of this control loop
was installed and worked reliably.
Without an alignment control of the cavity mirrors the 
continuous lock durations were limited by alignment drifts. However, we still
achieved continuous lock periods of up to 10 hours, and the automation
was able to relock the cavity over periods of up to 36 hours before 
the mirrors had
to be realigned. For comparison, the laser and mode cleaner
systems which were under automated alignment control \cite{hrg:aa} 
typically achieved continuous lock times of 48 hours. 
The stable operation of the entire system over periods of 36 hours showed
the stability of the involved subsystems, especially of the
seismic isolation and the control electronics.

\begin{figure}[htb]
\begin{center}
\IG [scale=1] {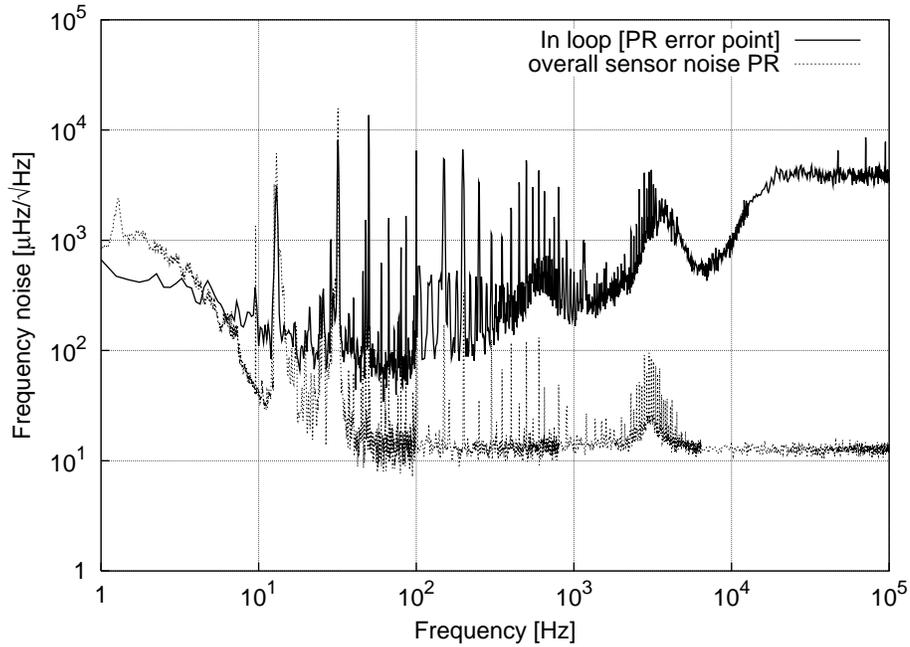} 
\end{center}
\caption{\label{fig:prc}
In-loop frequency noise of the control loop of the 1200\,m long cavity (PR loop):
about 100\,$\rm \mu Hz/ \sqrt{\rm Hz}$ at 100\,Hz.}
\end{figure}

For the GEO\,600 detector in its final state the required frequency
stability inside the power recycling cavity is 10\,$\rm \mu Hz/\sqrt{\rm Hz}$ at 100\,Hz,
which corresponds to a frequency stability of 100\,$\rm \mu Hz/\sqrt{\rm Hz}$ at
100\,Hz for the injected light.
For the intermediate experiment described here the goal was to achieve an in-loop frequency
noise at this level. Figure~\ref{fig:prc} shows the in-loop noise (i.e. the error point spectrum) of
the control loop for the 1200\,m long cavity and the sensor
noise of that loop. It can be seen that the goal mentioned above is met.

As of summer 2001, all main optics of the 
Michelson interferometer had been
installed and the power recycling cavity including a full 
interferometer is in place. This system has two longitudinal 
degrees of freedom, the
Michelson interferometer operating point (differential arm length) and the
power recycling length (common arm length plus the distance from the
power recycling mirror to the beam splitter). The error signals for each
degree of freedom depend strongly on the state of the other.
In order to lock the Michelson interferometer
the power recycling cavity is stabilised first. The experience with the 30\,m prototype
interferometer in Garching \cite{ghh:dr} showed that this locking hierarchy
works reliably. 
To compensate for the variable reflectivity of
the Michelson interferometer (when the interferomter is not yet
locked to the dark fringe), 
an automatic gain control has been added to the servo system 
for the power recycling cavity. The otherwise unchanged servo system 
can lock the power recycling cavity long enough for about 4 or 5 slow fringes
of the Michelson interferometer to pass.
During this time the Michelson interferometer error signal  
allows the lock acquisition of the Michelson interferometer.
Currently the servo system
for the Michelson interferometer control is tested. 
%Soon we will lock the power recycled Michelson 
%interferometer using the frequency stabilisation
%described in this paper.

\section{Conclusion}
The {\it 1200\,m experiment} described in this paper employed the first 
large-scale
optical system in GEO\,600. With the laser, the two mode cleaner systems 
and a 1200\,m
long cavity we could demonstrate the frequency stabilisation system of 
GEO\,600.
The required frequency noise reduction was achieved by stabilising the laser 
only to suspended
cavities.
The frequency and length control servos are automated and controlled by a 
computer system. 
The described system 
can acquire lock reliably and stay locked for 10 hour long periods. 
We demonstrated that the system works as expected so that 
the requirement for the laser frequency noise in GEO\,600 gravitational
wave detector can be met.
 
\ack
The authors would like to thank the British Particle Physics and 
Astronomy Research Council (PPARC), the German Bundesministerium 
f\"ur Bildung und Forschung (BMBF) and the State of Lower Saxony (Germany).

\end{document}